\documentclass[12pt]{article}
\usepackage{epsfig}
\usepackage{amssymb}

\newcommand{\be}{\begin{equation}}
\newcommand{\ee}{\end{equation}}
\newcommand{\bea}{\begin{eqnarray}}
\newcommand{\eea}{\end{eqnarray}}
\newcommand{\ov}{\overline}
\newcommand{\ve}{\varepsilon}

\newcommand{\ba}{\begin{array}}
\newcommand{\ea}{\end{array}}

\def\fun#1#2{\lower3.6pt\vbox{\baselineskip0pt\lineskip.9pt
\ialign{$\mathsurround=0pt#1\hfil##\hfil$\crcr#2\crcr\sim\crcr}}}

\title{Formation of light antinuclei and "dense gas" stage in heavy ion
 collisions}
\author{I.A. Shushpanov \\
{\it \normalsize Institute of Theoretical and Experimental
Physics},\\ {\it \normalsize B.Cheremushkinskaya 25, 117218
Moscow,Russia}\\ {\tt \normalsize shushpan@heron.itep.ru}}
\date{}

\begin{document}
\maketitle

\vspace{10mm}

\begin{abstract}
 The antideuteron and antihelium-3 production rates at high-energy
 heavy ion collisions are calculated in the framework of fusion mechanism. It is
 supposed, that $\bar p$,$\bar n$,$\bar d$,${}^3 \ov He$ participating in the fusion are moving in the
 mean field of other fireball constituents. It is demonstrated that at the
 "dense gas" stage of fireball evolution,
 preceding the thermal freeze-out, the coalescence parameters for $\bar d$ and
 ${}^3 \ov He$ can be found from the requirement of balance between created and
 disintegrated antinuclei. The explicit formulae for coalescence parameters are
 presented and compared with the data.

\end{abstract}

\section{Introduction}

In recent experiments on heavy ion collisions the production of
 light antinuclei was measured \cite{1,2,3,4}. One may expect that this
 process proceeds at the intermediate stage of the evolution of
 the fireball, created in heavy ion collisions. Because of their
 small binding energy antinuclei are formed at the stage when the
 hadrons are already formed, the density of hadronic matter is of
 the order of normal nuclear density, and the particle collisions are
 still important. We will call this stage as the "dense gas"
 stage of fireball evolution.
 It is a common belief, that the antinuclei production proceeds as a
 fusion process:
 $\bar{p} \, + \, \bar{n}\,  \to \, \bar{d}$
 in case of antideuterons and
$\bar{p}\,+\,\bar{p}\,+\,\bar{n} \, \to \, ^3 \ov{He}$ in case of
$^3\ov{He}$. According to the dominant coalescence mechanism it is
convenient to characterize $\bar{d}$ and $^3\ov{He}$ production by
the coalescence parameters
\begin{equation}
B_2\, =\,E_{\bar{d}} \,{d^3N_{\bar{d}} \over d^3p_{\bar{d}}}
 \left(  E_{\bar{p}} {d^3N_{\bar{p}} \over d^3p_{\bar{p}}}
 \right)^{-2}; \qquad B_3\, =\,E_{\ov{He}}\, {d^3 N_{\ov{He}} \over
d^3p_{\overline{He}}} \left( E_{\bar{p}} {d^3N_{\bar{p}} \over
d^3p_{\bar{p}}} \right)^{-3} , \label{3} \end{equation} where one
assumes $d^3N_{\bar{n}}/d^3p_{\bar
n}=d^3N_{\bar{p}}/d^3p_{\bar{p}}$.

The basic ideas  of our approach are the following. We assume,
that the coalescence mechanism is the dominant one in the
production of heavy nuclei. The fusion reactions cannot proceed if
all particles are on
 mass shell. However, in the fireball at the dense gas stage
 of its evolution,
$\bar{p},\bar{n},\bar{d},~^3\ov{He}$ are not on mass shell, since
they interact with surrounding matter. One may consider their
movement as a propagation in the mean external complex field
caused by the matter. The interaction with this field leads to
appearance of the mass shifts and the widths of all particles
propagating in the medium (or width broadening for unstable ones),
analogous to refraction and attenuation indices in the case of
photon propagation. Another important ingredient of our approach
is the balance of antinucleous production and disintegration
rates. This balance is achieved because of large density of pions
in the fireball and high rate of $\pi +$ ($\bar{d}$ or
$^3\ov{He}$) collisions leading to antinucleous disintegration.
The balance does not imply a statistical equilibrium, but rather a
stationary process. The statistical or thermal equilibrium are not
assumed in the calculation. This work is based on the results
obtained in \cite{d,He}.

\section{Calculation of the coalescence parameters}
Let us consider first the $\bar{d}$-production. We will use the
notation $q_i(x,p)$, $i=\bar{p},\bar{n},\bar{d},\pi, ...$ for the
double densities in coordinate and momentum spaces and
$n_i(x)=\int q_i(x,p)\, d^3p$ for the coordinate densities.
($q_i(x,p)$ are Lorentz invariant.) Let us choose the c.m.~frame
of colliding ions. The transport equation for $q_{\bar{d}}(x,p)$
has the form: \bea {m_d\over E_{\bar{d}}} \, {\partial
q_{\bar{d}}(p_{\bar{d}},x)\over\partial
x_{\mu}}u^{\bar{d}}_{\mu}\,
 =\, {\partial q_{\bar{d}}\over \partial t}\, +\, {\bf v}_{\bar{d}}
\bigtriangledown q_{\bar{d}}\, = \nonumber\\
 =\, \int d^3
p_{\bar{p}}\, d^3 p_{\bar{n}}\, q_{\bar{n}}(p_{\bar{p}})
 \, q_{\bar{n}}(p_{\bar{n}}) \, \omega_{\bar{p}\bar{n}\to \bar{d}} \,
-\,q_{\bar d}(p_{\bar d}) \sum_i \int d^3 p_i \, q_i(p_i) \,
\omega_{{\bar d}i\to X} \label{20} \eea where
$u^{\bar{d}}_{\mu}=(1,{\bf v}_{\bar{d}})/\sqrt{1-v^2_{\bar{d}}}$
is antideuteron 4-velocity, $\omega_{\bar{p}\bar{n}\to \bar{d}}$
is the fusion reaction
 propability proportional  to differential cross section which can
 be written at low (nonrelativistic) c.m. energies as
\be
\omega_{{\bar p}{\bar n}\to {\bar d}} \, =\,{3\pi g^2\over 16
E_{\bar p} E_{\bar n} E_{\bar d}} \ \delta^4(p_{\bar p}+p_{\bar
n}-p_{\bar d}), \label{omsig} \ee ($E_{\bar p}$ and $m$ are
antiproton energy and mass, etc.) and similarly for the
desintegration process ${\bar d}i\to X$ due to collisions of
$\bar{d}$ with $i$-th constituent of the fireball ($i=\pi,K,p,n$,
etc.).
 Low energy coupling
constant $g^2$ was found by Landau many years ago \cite{10} and
with account of the correction due to nonzero radius of nuclear
force becomes equal to ($(1-\alpha r_0)^{-1}\approx 1.67$)
\be
g^2\, =\, 128 \pi\, m\sqrt{m\ve}\,(1-\alpha r_0)^{-1} \label{9}
\ee

At the dense gas stage of the fireball evolution all particles
inside the fireball should be considered as moving in the mean
field of other fireball constituents. As a consequence, the masses
 are shifted in comparison with their vacuum values. Similarly,
 due to interaction with medium constituents, the widths
 $\Gamma_i$ appear. The mass shift $\Delta m(E)$ and the width
 $\Gamma(E)$ are expressed in terms of the forward scattering
 amplitude $f(E)$ of the particle on $i$-th medium constituents
 (see\cite{15,16} and references therein)
 \be
 \Delta m(E)\, =\, -2\pi\, {n_i\over m}\, {\rm Re}\,f_i(E)\,;  \qquad \qquad
 \Gamma(E)\, = \, 4\pi\, {n_i\over m}\, {\rm Im} f_i(E)\,=\,{n_i p\over m}\, \sigma_i(E), \label{22}
\ee where $E,p$ and $m$ are particle energy, momentum and mass,
$n_i$ is the density of $i$-th constituent in medium.
Eq's.(\ref{22}) take place in the system, where the constituents
are at rest.

Therefore, $\bar{p},\bar{n}$ and $\bar{d}$ can be considered as
Breit-Wigner resonances with varying  masses distributed according
to Breit-Wigner formula. In the process of the fireball  expansion
these Breit-Wigner resonances smoothly evolve to their stable
counterparts. So, we substitute ($\ref{omsig}$) in the first term
in the right-hand side of ($\ref{20}$) and integrate over the
masses of Breit-Wigner resonances. The calculation gives the
following result for production term in the transport equation
\be
 I \, =\, {3 \pi^2\over {8 E_{\bar d}}} \,  g^2 \,
\sqrt{ \Gamma (1+a) \over 2 m} \, q^2_{\bar{p}} (p_{\bar{p}})
\label{24}
 \ee
Here $\Gamma_{\bar p}=\Gamma_{\bar n}\equiv\Gamma$ and parameter
$a \sim 1$ describes the probability of $\bar d$ production from
the system with $\bar{d}$ quantum numbers.

 The calculation of the second (disintegration) term in ($\ref{20}$) is very simple
 as $\Gamma_{\bar d}$ has the meaning of $\bar d$ decay rate in the
 rest system. Performing Lorentz boost, we obtain this term at arbitrary
 reference frame
 \be
 II=q_{\bar{d}}(p_{\bar d}){m_d\over E_{\bar d}}\,\Gamma_{\bar d}
 \ee

 The balance condition determines $\bar{d}$-density
 and therefore $\bar d$ coalescence parameter
\be
B^{th}_2 \, =\, {24 \pi^3\over E_{\bar{p}}} \times 1.67\,
\sqrt{\ve(1+a)\over 2\Gamma} \, {2\over V}\,
{\ov{n^2}_{\bar{p}}\over (\bar{n}_{\bar{p}})^2}, \label{33} \ee
where $V$ is the fireball volume, $\bar{n}_p$ and $\ov{n^2}_p$ are
the mean and mean square $\bar{p}$ densities in the fireball. An
additional factor 2 in $B^{th}_2$ appears because at given
annihilation length only half of particles reaches a detector.

The calculation of $^3\ov{He}$ production proceeds along the same
lines as $\bar{d}$. In this case the value of low energy coupling
constant can be estimated as (see \cite{He} for details)
\be
G^2\, =\, 36 \sqrt{3}\,(4\pi)^3 \, {m\over \Lambda} \label{19} \ee
In numerical calculations $\Lambda=300\,{\rm MeV}$ will be taken.

The result for ${}^3 \ov He$ coalescence parameter is
\be
B_3\, =\,96\pi^7\, {1\over \sqrt{2}} \, {\Gamma\over \Lambda}\,
{1\over V^2} \,{\ov{n^3}\over \bar{n}^3} \frac{1}{E^2_{\bar p
}}\label{38} \ee

\section{Comparison with experimental data}
 We accept the following model for the dense gas stage of fireball
 evolution \cite{16}. Assume that  any participant -- nucleon or pion
 occupies the volume $v_N=1/n^0_N$ or $v_{\pi}$, respectively.
 For numerical estimations we take $n^0_N = 0.24 \, {\rm fm}^{-3}$
 and $\beta = v_{\pi}/v_N=(r_{\pi}/r_N)^3 \approx 0.55$,
 where $r_{\pi} = 0.66\, {\rm  fm}$
 and $r_N = 0.81 \, {\rm fm}$ are pion and nucleon electric radii.
 From estimation of antiproton annihilation length we can put with sufficient
 accuracy $\ov{n^2}/\bar{n}^2\approx 2$ and $a=1/2$.
 The results for $\bar d$ coalescence parameter are collected in the
 Table~\ref{table:1}.

\begin{table}[htb]
\caption{Comparison with experimental data. } \label{table:1}
\newcommand{\m}{\hphantom{$-$}}
\newcommand{\cc}[1]{\multicolumn{1}{c}{#1}}
\renewcommand{\tabcolsep}{2pc} 
\renewcommand{\arraystretch}{1.2} 
\begin{tabular}{@{}llll}
\hline Experiment           & \cc{NA44} & \cc{STAR} & \cc{E684}
\\ \hline
$B^{exp}_2$ ($10^{-4}\, {\rm GeV}^2$)& \m $4.4\pm 1.3$ & \m
$4.5\pm 0.3\pm 1.0$ & \m $41\pm 29\pm 23$ \\ $B^{th}_2$
($10^{-4}\,{\rm GeV}^2$) & \m 4.0 & \m 4.2 & \m 14.0 \\ $\Gamma$
(MeV) & \m190 & \m210 & \m140 \\ $V$ ($10^3\,{\rm fm}^3$) & \m 6.0
& \m 7.8 & \m 3.0
\\ \hline
\end{tabular}\\[2pt]
\end{table}

 For antihelium STAR found:
\be
B^{exp}_{3, \, \ov{He}}\,=\,(2.1 \pm 0.6 \pm 0.6)\times
10^{-7}\,{\rm GeV}^4 \label{39} \ee The theoretical value for
$B^{th}_3$ at $n^0_N=0.24\,{\rm fm}^{-3}$, $\ov{n^3}/\bar{n}^3=3$,
and $\Lambda=300\,{\rm MeV}$ is:
\be
B_{3,\,\ov{He}} \,=\,3.3\times 10^{-7}\,{\rm GeV}^4 \label{40} \ee
The agreement with experiment is good, despite of many theoretical
uncertainties. It demonstrates the validity of basic ideas of
theoretical approach.

\section{Summary and Acknowledgements}
The coalescence parameters for production of antideutrons and
antihelium-3 in heavy ion collisions were calculated. The obtained
results are based on three assumptions: i) the  main mechanism of
light antinucleous  production is coalescence (fusion) mechanism;
ii) all particles, participating in fusion  process are moving in
the mean field of other fireball constituents; iii) the number of
produced antinucleous is determined by the balance conditions: the
equality of produced and disintegrated --- mainly by pions ---
antinuclei.  The production of antinucleous proceeds at the dense
gas stage of the fireball evolution, when the hadrons are already
formed, but their interactions are still important. Statistical or
thermal equilibrium are not supposed at the dense gas stage. In
fact, the final results depend on one parameter --- the volume of
the fireball at this stage (or, equivalently, on the hadron
densities.) Good agreement with experimental data for coalescence
parameters was obtained for experiments at CERN, RHIC and AGS for
the values of $n^0_N\sim 0.17 \div 0.30\,{\rm fm}^{-3}$, close to
normal nucleous density. Much lower, or much higher values of
$n^0_N$ lead the to values of coalescence parameters, incompatible
with the data.

\vspace{5mm }

This work was supported in part by INTAS grant 2000-587 and RFBR
grant 03-02-16209.

\end{document}